\documentclass[aip,chaos,preprint]{revtex4-1}

\usepackage{amsmath,amssymb}

\usepackage{graphicx}
\usepackage{xcolor}
\usepackage{hyperref}
\usepackage{mathrsfs}

\begin{document}

\title{Emergence of Statistical Financial Factors by a Diffusion Process}

\author{Jose Negrete Jr}
\affiliation{Instituto de Ciencias Sociales y Administraci\'on (ICSA),\\
Universidad Aut\'onoma de Ciudad Ju\'arez (UACJ), Ciudad Ju\'arez, Chihuahua, M\'exico.}
\email[Corresponding author: ]{jose.negrete@uacj.mx}

\author{Jaime Joel Ramos}
\affiliation{School of Business, University of North Texas at Dallas (UNTD),\\Dallas, Texas, USA.}

\begin{abstract}
\noindent Factor models provide a representation of the joint behavior of large cross-sections of financial assets through a reduced set of underlying drivers. In this work, we propose a network-based dynamical framework in which statistical factors emerge endogenously from interactions among assets, rather than being imposed exogenously or extracted purely statistically. We model asset returns as a system of coupled nonlinear maps influenced by a network of interactions encoded in a coupling matrix. This matrix is constructed via an orthogonal transformation of a Laplacian operator with prescribed nullity, allowing us to control over the system's effective dimensionality. Under appropriate coupling conditions, the dynamics reduce from a high-dimensional phase space to a lower-dimensional invariant manifold, where synchronization modes of co-movement arise. We show that the number of emergent statistical factors is directly related to the nullity of the coupling matrix, consistent with a stability analysis performed here. Simulations demonstrate that the model reproduces key stylized features of financial returns while generating factors with balanced loadings across assets. Finally we show that this model reproduces features that are observed in an empirical portfolio. These findings suggest that statistical factors can be interpreted as emergent synchronization modes of an interacting nonlinear system, providing a complementary perspective to existing economic and statistical factor models.
\end{abstract}


\maketitle

\noindent\textbf{Statistical factors correspond to a reduced number of signals that describes the time evolution of a large cross-section of stocks. Here we wonder what their possible origin might be. One possibility is that stocks reflect the movements dictated by exogenous signals. An opposing view is that traders make their decisions given the previous movements of a stock i.e. behaving as a dynamical system. We use a combination of dynamical systems and network theory to propose that factors might emerge from the activity given by traders.}

\section{Introduction}
Factor models play a central role in empirical finance, providing a representation of the joint behavior of large cross-sections of asset returns through a reduced set of underlying drivers. These underlying drivers determine the systemic risk behind an investment portfolio, i.e. the noise that cannot be mitigated by diversification~\cite{malkiel2023random}.  In these models, the return of $K$ assets,  denoted by the vector $r_t \in \mathbb{R}^K$ 
are expressed as a linear combination of a smaller set of latent factors $f_t \in \mathbb{R}^M$, 
with $M\ll K$:
\begin{equation}
	r_t= \mathbf{B} f_t+\xi_t.
\label{e1}
\end{equation}
	
\noindent where $\mathbf{B} \in \mathbb{R}^{K \times M}$ is the loading matrix, whose elements $\beta_{k,m}$ quantify the exposure of each stock to a given factor, and $\xi_t \in \mathbb{R}^K$ captures deviations (or noise) not explained by the factors. 

Factor models can be broadly categorized into economic factors, motivated by financial theory, and statistical factors, inferred directly from data. The Capital Asset Pricing Model (CAPM) is the first economic factor model proposed \citep{sharpe1964capital}, in which returns depend on a single market-wide factor. This model gives a mathematical definition for the risk premium, which is the excess return a stock has as a function of risk. However, independent tests with empirical data does not fulfill the CAPM predictions between risk and return \citep{jensen1972capital}. Subsequent extensions, such as Fama-French three-factor and five-factor models, incorporate firm characteristics such as size, value, and profitability.\citep{fama2004capital, fama2015fivefactor}.

In contrast, statistical factor models use techniques such as principal component analysis (PCA) to extract orthogonal components that explain the covariance structure of returns without requiring an explicit economic interpretation \citep{chamberlain1982arbitrage, connor1993test}. Several authors have used random matrix theory to determine the number $M$ of factors behind a given cross section of stocks \citep{laloux2000random, plerou1999universal, plerou2002random}. Interestingly, the CAPM market financial factor displays a strong correlation with the first component extracted via PCA \citep{avellaneda2022principal}. These methods are widely used due to their flexibility and empirical performance.

Despite their success, a fundamental question remains: why do statistical factors emerge in the first place? As discussed, most approaches treat factors either as exogenous drivers derived from macroeconomic variables or as statistical constructs obtained from dimensionality reduction. However, these perspectives do not provide a structural explanation for the origin of factors.

Under the efficient market hypothesis (EMH), stock prices reflect all available information including both public and private. Because the information set influencing prices is vast, traditional fundamentals (such as dividends and earnings) represent only one component of a much broader information set. Thus, according to EMH, the stock prices are driven by random exogenous signals defined by all the available information in the economy~\cite{malkiel2023random}. At the same time, empirical evidence suggests that financial markets exhibit deviations from purely information-driven dynamics, including herding behavior, feedback trading, and endogenous fluctuations~\cite{sornette2015financial}. These effects introduce interactions between assets, including that between investors and the market itself, that can be modeled using dynamical systems. Several works have explored the effect of different feedback mechanisms such as trend following~\cite{palmer1994artificial}, adaptation of expectations~\cite{arthur1996asset, brock1997rational} and leverage~\cite{thurner2012leverage}. This observation motivates an alternative perspective: rather than imposing factors externally or extracting them purely statistically, we investigate whether factors can emerge endogenously from the interactions among assets. Modern financial markets are highly interconnected through shared information channels, overlapping investor bases, sectoral linkages, and institutional flows. If asset returns are coupled through such interactions, coherent patterns of co-movement may arise spontaneously as collective modes of the dynamical system.

From the viewpoint of nonlinear dynamics, this phenomenon can be interpreted as a dimensionality reduction mechanism. Matrices can be used to define the structure of the interconnections between stocks.  In particular, previous work has shown that high-dimensional systems can exhibit effective low-dimensional behavior of self-organization in finance \citep{borland2012statistical, bouchaud2024selforganized}, as transitions to a highly coordinated states were observed in both the dot-com and housing crisis. Dimensionality reduction can be achieved by the emergence of synchronization modes~\cite{pikovsky2001universal}.

In this paper, we propose a network-based dynamical model in which statistical factors arise as emergent modes of a system of interacting assets. We model asset returns as a set of coupled nonlinear maps connected through a network structure encoded by a coupling matrix. We exploit synchronization theory for chaotic systems~\cite{pikovsky2001universal} and show that, under appropriate conditions, the system exhibits a reduction from $K$ effective degrees of freedom to $M$ dominant modes, which formalize this structure and can be interpreted as statistical and potentially as financial factors. In this context, interactions among assets propagate through the coupling network in a diffusion-like manner; accordingly, the term diffusion is used throughout this paper in the sense of the propagation of dynamical influence across a network of interacting assets, rather than in the stochastic-process sense commonly used in mathematical finance. Our contribution is therefore a generative mechanism that explains how statistical factors may arise from endogenous interactions. To avoid ambiguity, we distinguish between synchronization modes, which are collective dynamical structures of the coupled system; statistical factors, which are latent components extracted from returns using methods such as PCA; and financial factors, which are economically motivated factors such as those appearing in CAPM and Fama–French models. This perspective complements traditional approaches and existing models by linking the emergence of statistical factors to the underlying dynamics of interconnected financial systems.

\section{The Model}
\subsection{A Chaotic Model for Asset Returns}

Motivated by inter-day returns time series typically showing an absence of autocorrelation, heavy-tailed distributions \citep{cont2001empirical}, and evidence that financial time series evolve with some degree of chaotic dynamics\citep{holyst2001observations,krishnadas2022recurrence}, we consider a system of $K$ interacting assets whose return vector at time $t$ is denoted by $r_t \in \mathbb{R}^K$ with the following characteristics. 

The temporal evolution of returns is modeled using a system of coupled nonlinear maps: 
\begin{equation}
	r_{t+1}=\left(1-\varepsilon\right)g\left(r_t\right)+\frac{\varepsilon}{N} \mathbf{C} g\left(r_t\right),
\label{e2}
\end{equation}
\noindent where $g(\cdot)$ is a nonlinear transformation applied component-wise, $\mathbf{C} \in \mathbb{R}^{K\ \times K}$ is a coupling matrix, and $\varepsilon \in [0, 1]$ controls the strength of interactions among assets. 
The coupling matrix $C$ is a connectivity matrix that determines how the $K$ components of $r_t$ influence each other. The coupling parameter $\varepsilon$ controls the components' degree of independence, thus $\varepsilon=\ 0$ indicates that each asset evolves independently according to 
\begin{equation}
r_{t+1}^{\left(k\right)}=g\left(r_t^{\left(k\right)}\right),
\label{newe1}
\end{equation} whereas for $\varepsilon > 0$, assets become coupled through the network defined by $C$. In the limit $\varepsilon=\ 1$, the dynamics are fully driven by interactions among assets.

\subsection{The Local Map $g\left(r_t\right)$}

The nonlinear map $g(\cdot)$ is constructed to reproduce key stylized facts of financial returns, including heavy-tailed distributions and weak autocorrelation. The construction is defined by the following three steps:

\begin{equation}
		u_t=h\left(r_t\right)
\label{e3}
\end{equation}

\begin{equation}
	u_{t+1}=\frac{u_t}{\delta}-\left\lfloor\frac{u_t}{\delta}\right\rfloor	
\label{e4}
\end{equation}

\begin{equation}
	r_{t+1}=h^{-1}\left(u_{t+1}\right),
\label{e5}
\end{equation}

\noindent where $u_t \in [0,1]$ evolves according to a Bernoulli map,   a canonical source of chaotic dynamics \citep{ott2002chaos}. Moreover, for $\delta$ small enough, the variable $u_t$ is uniformly distributed with  probability density function (pdf) $\rho\left(u_t\right) = U\left[0,1\right]$. This allows full control of the pdf of $r_t$ when $\varepsilon=\ 0$, since 
\begin{equation}
\rho\left(r_t\right)=\frac{dh\left(r_t\right)}{dr_t}.
\end{equation}

\noindent In general, the transformation $h(\cdot)$ is chosen to satistfy any desired properties. Thus, it is chosen so that the resulting stationary distribution of returns follows a heavy-tailed density: 
\begin{equation}
	\rho\left(r_t\right)=\frac{\gamma}{\pi}sech\left(\gamma\left(r_t-r_0\right)\right),
\label{e6}
\end{equation}

\noindent with mean $\mu=r_0$, variance $\sigma^2=\pi^2/4\gamma^2$ and excess kurtosis $\kappa=\ 2$.

\begin{figure}[tb]
\centerline{\includegraphics[width=5in]{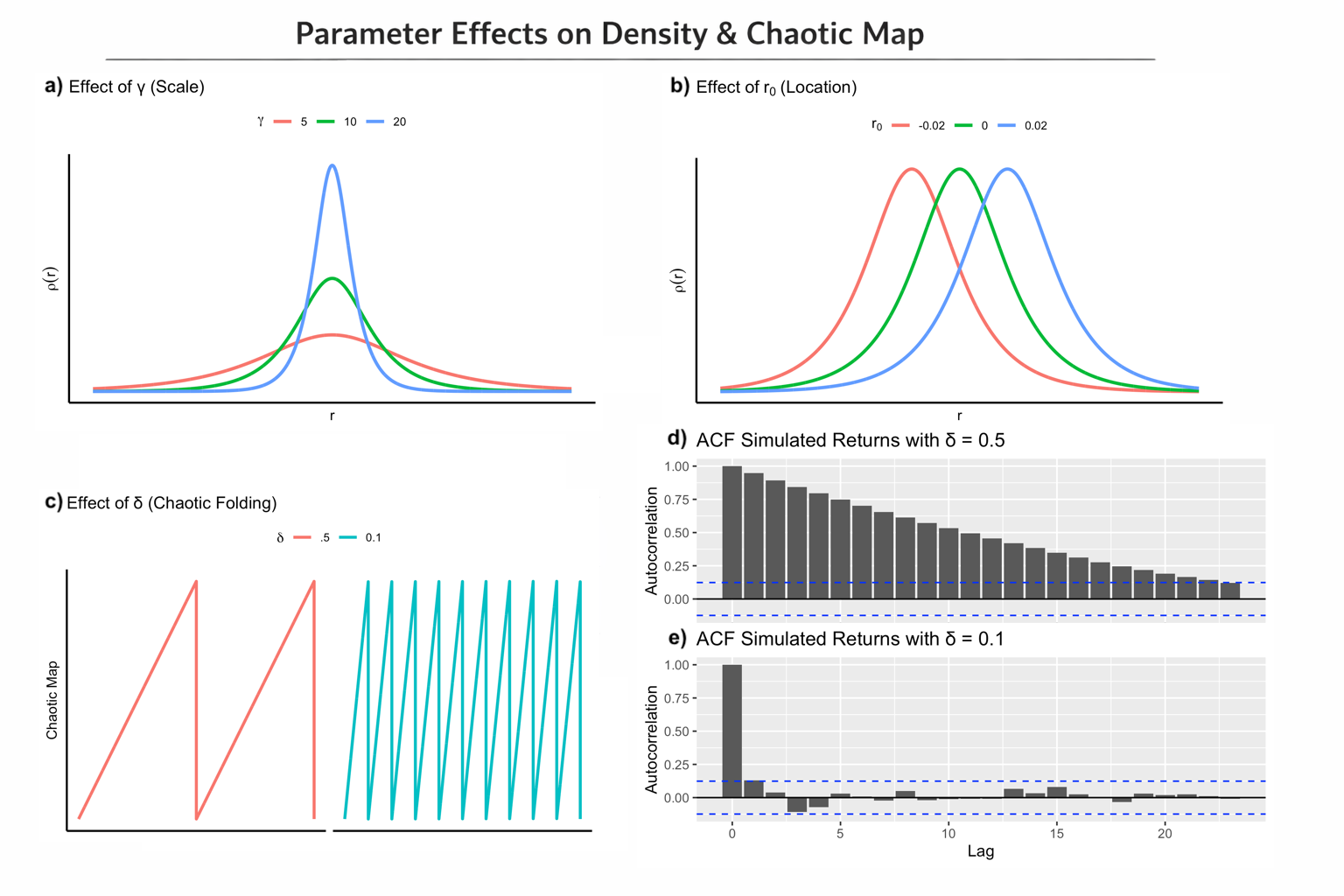}}
\vspace*{8pt}
\caption{Effects of model parameters on the distribution and dynamical properties of simulated returns. 
(a) Probability density function defined in Eq. (\ref{e6}) for different values of $\gamma$, showing that larger $\gamma$ reduces the variance $\sigma^2=\pi^2/4\gamma^2$.
(b) Effect of the location parameter $\mu=r_0$, shifting the distribution from along the return axis.
(c) Chaotic map defined by Eq. (\ref{e4}) illustrating the influence of $\delta$, where larger values of $\delta$ produces less noisy series when compared to small values producing rapid changes. 
(d)--(e) Simulated return series for $\gamma = 60$ and $r_0 =0.001$ with different $\delta$: (d) $\delta = 0.5$ exhibits strong autocorrelation, while (e) $\delta=0.1$ shows negligible autocorrelation, as supported by Ljung--Box tests and Durbin--Watson statistics. This illustrates how $\delta$ modifies the chaotic dynamics, governs autocorrelation properties, and allows calibration to match empirical stylized facts.}
\label{f1}
\end{figure}

Furthermore, this construction enables control over the statistical properties of the generated time series. In particular, suitable choices of the parameter $\delta$ produce series of returns with negligible autocorrelation, while preserving large fluctuations consistent with empirical financial data. Fig. \ref{f1} illustrates the wide tail distribution from Eq. (\ref{e6}), along with the effect of initial parameters $\gamma$ (Fig. \ref{f1} (a)) and $r_0$ (Fig. \ref{f1} (b)). Fig. \ref{f1} (c-d) shows the influence of the parameter $\delta$ in the chaotic map of Eqs. (\ref{e3})-(\ref{e5}), where smaller $\delta$ generate faster fluctuation and negligible autocorrelation; larger $\delta$ values yield smoother auto correlated time series.

\subsection{Empirical Relevance of Local Dynamics}

To validate the empirical relevance of the local dynamics, we performed a grid search over parameters $(r_0, \gamma, \delta)$ and compared simulated returns with observed returns from a representative stock. We computed the mean square error: \begin{equation}
	\mathrm{MSE}\ =\ \frac{1}{T}\sum_{t=1}^{T}\left(r_t\ -\ R_t\right)^2,	
\label{e7}
\end{equation}

\noindent between the generated series $r$ and the Netflix (NFLX) return series $R$ over $T=251$ trading days (25/07/2022 - 25/07/2023). Using a fixed random seed, we generated returns over a parameter grid consisting of:
$$r_0 \in seq(-0.02, 0.02, by = 0.001) \quad \gamma \in seq(5, 100, by = 1)$$ $$\delta \in seq(0.005, 0.02, by = 0.001) \cup c(0.25, 0.03, 0.04, 0.05, 0.75, 0.1).$$
Each resulting series was evaluated using the MSE, the Ljung-Box test ($H_0:$ no autocorrelation), and the Durbin-Watson statistic (DW) which jointly provide insight into autocorrelation and overall fit.

\begin{figure}[bb]
\centerline{\includegraphics[width=6.5in]{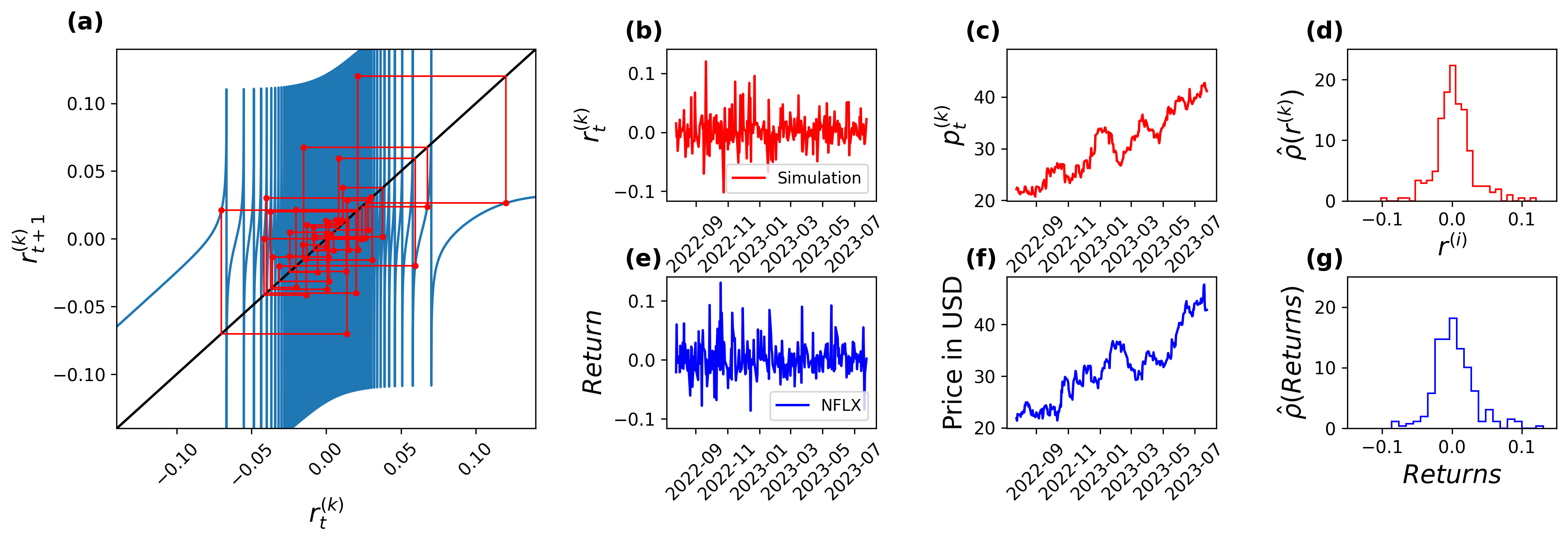}}
\vspace*{8pt}
\caption{Comparison between simulated dynamics and empirical financial data.
(a) Iterative nonlinear map $r_{t+1}^{\left(k\right)}=g\left(r_t^{(k)}\right)$ with cobweb trajectory. 
(b)--(d) Simulated return series, corresponding price evaluation, and estimated probability density function for a single realization.
(e)--(g) Empirical return series (NFLX), corresponding price trajectory, and estimated density for the period 25/07/2022 - 25/07/2023. The model reproduces key empirical features such as heavy tails and intermittent large returns fluctuations.}
\label{f2}
\end{figure}

A total of 86,591 time series were generated, demonstrating that the model can reproduce realistic statistical properties across a wide range of parameters. Approximately 15\% produced unstable series with less than 251 observations. Among the 73,438 fully generated series, 75\% achieved an MSE $\leq 0.0042$, indicating that the model can match empirical returns with high accuracy across a broad range of initial parameters. Low values of $\delta$ generally yielded negligible autocorrelation, consistent with observations on stock returns \citep{cont2001empirical}. Focusing on $\delta$, roughly 86\% of the complete series had a Ljung-Box $p-$value $\geq  0.05$ and a DW statistic in the interval $(1.75, 2.25)$, both of which indicate the absence of autocorrelation. Specifically, $\delta$ values of $0.005, 0.010, 0.020$ and $0.050$ produce tended to produce series with MSE $> 0.005$ or incomplete trajectories. These conclusions held robustly across all values of $r_0$. Accordingly, and without loss of generality, the remainder of this work adopts the parameter values $\delta= 0.011$, $\gamma= 60$ and $r_0=0.001$. These are the values used in Fig. \ref{f2} (b), accurately representing NFLX (Fig. \ref{f2} (e)) with a MSE of $0.0016$.

Fig. \ref{f2} (a) displays the mapping $r_{t+1}^{\left(k\right)}=g\left(r_t^{\left(k\right)}\right)$ for $\varepsilon=\ 0$ along with its cobweb diagram. (b) - (d) display the associated time trace of the return $r_t^{\left(k\right)}$, the associated price series $p_{t+1}^{\left(k\right)}=\left(1+r_{t+1}^{\left(k\right)}\right)\ \ p_t^{\left(k\right)}$, and the estimated pdf $\hat{\rho}\left(r_t^{\left(k\right)}\right)$. For comparison, (e) - (g) present the same quantities for NFLX. It is significant to highlight how simulated traces exhibit NFLX behavior, including sudden large jumps, which manifest as heavy tails in the estimated pdf. Our model produces these jumps endogenously rather than being driven by external stress, consistent with studies showing that large return movements are intrinsic to asset dynamics rather than triggered by external news \citep{joulin2008stock}. 

\begin{figure}[tb]
\centerline{\includegraphics[width=6.5in]{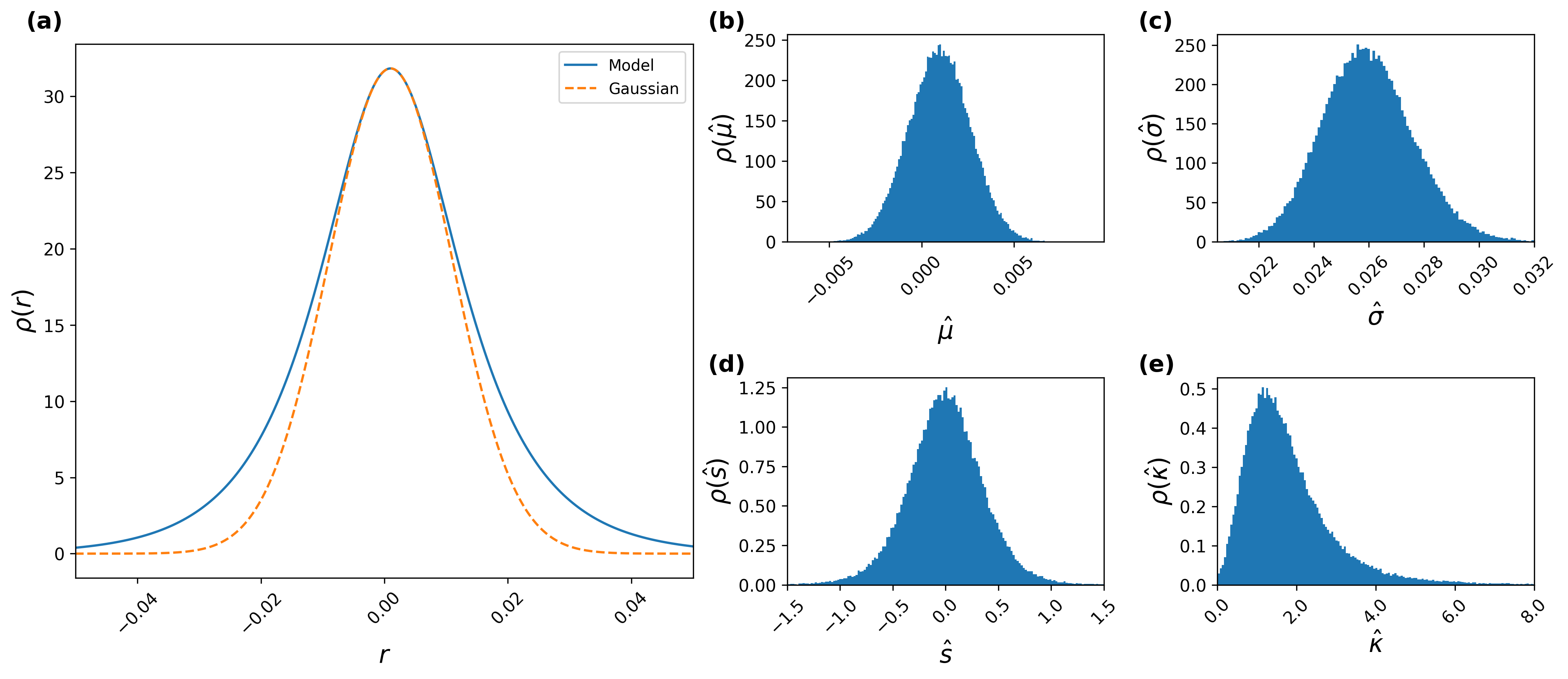}}
\vspace*{8pt}
\caption{Statistical properties of simulated returns compared with a Gaussian benchmark. (a) Probability density function $\rho(r)$ from Eq. (\ref{e6}) (solid line) compared to a Gaussian distribution (dashed line), highlighting heavier tails and excess kurtosis $\kappa=\ 2$. 
(b)--(e) Sampling distributions of the mean $\hat{\mu}$, standard deviation $\hat{\sigma}$, skewness $\hat{s}$, and kurtosis $\hat{\kappa}$ computed from 100,000 simulated observations of Eq. (\ref{newe1}), showing broader variability than in the Gaussian case.} 
\label{f3}
\end{figure}

Fig. \ref{f3} (a) further compares the model-implied heavy-tail density produced by Eq. (\ref{e6}) using randomly drawn $r_0$ and, as previously stated, fixed $\delta= 0.011$ and $\gamma=60$, over 100,000 iterations with a Gaussian distribution benchmark. With the chosen parameters, the resulting pdf $\rho\left(r_t\right)$ has a mean of $\mu= 0.001$ and a standard deviation of $\sigma= 0.026$ (Fig. \ref{f3} (a)), a relationship consistent with empirical financial data in which the standard deviation exceeds the mean by an order of magnitude \citep{bouchaud2003theory}. As a result of the slow decay in the tails, the convergence to the population mean is slower than $\sim1/\sqrt T$ \citep{taleb2020statistical}. The sampling distributions of mean $\hat{\mu}$, standard deviation $\hat{\sigma}$, skewness $\hat{s}$, and kurtosis $\hat{\kappa}$ (Fig. \ref{f3} (b) -(e)) are also wider than those expected for a normal distribution. Essentially, our model produces a universe of returns with realistic levels of variability and heavy-tailed behavior, though the variability is lower than that observed in actual markets\citep{bouchaud2003theory}. 

\subsection{Network Structure and Coupling Matrix}

The interaction structure among assets is encoded in the coupling matrix $\mathbf{C} \in \mathbb{R}^{K \times K}$. We construct $\mathbf{C}$ by first defining a Laplacian matrix $\mathbf{L}$ representing a system of $M$ initially disconnected clusters:
\begin{equation}
	\mathbf{L}=\left[\begin{matrix} \mathbf{L}_1&0&\cdots&0\\0& \mathbf{L}_2& \cdots&0\\\vdots&\vdots&\ddots&\vdots\\0&0&\cdots& \mathbf{L}_M\\\end{matrix}\right]	
\label{e8}
\end{equation}

\noindent where each block $\mathbf{L}_i \in \mathbb{R}^{N \times N}$ corresponds to a fully connected cluster of $N$ assets (Fig. \ref{f4} (a)):
\begin{equation}
	\mathbf{L}_i=\left[\begin{matrix}1-N&1&\cdots&1\\1&1-N&\cdots&1\\\vdots&\vdots&\ddots&\vdots\\1&1&\cdots&1-N\\\end{matrix}\right].	
\label{e9}
\end{equation}

The full matrix $\mathbf{L}$ is negative-semidefinite with nullity $M$ ($\lambda = 0$) and rank $K-M$ ($\lambda = -N)$. These properties play a center role in determining the effective dimension of the dynamics. To generate a fully interconnected network while preserving the spectral properties of $\mathbf{L}$, we apply an orthogonal transformation, a random rotation, defined:
\begin{equation}
	\mathbf{C}=\mathbf{QLQ}^T,
\label{e10}
\end{equation}	

\noindent where $\mathbf{Q} \in \mathbb{R}^{K \times K} \ (K = MN)$ is an orthogonal matrix from a $QR-$decomposition of a random matrix $\mathbf{A} \in \mathbb{R}^{K \times K}$ whose entries are drawn independently by the standard normal distribution $a_{i,j}\sim\mathcal{N}\left(0,1\right)$ and
\begin{equation}
\mathbf{A}\ =\ \mathbf{QR}.
\label{newe3}
\end{equation}

\noindent This transformation preserves the eigenvalues of $\mathbf{L}$ but removes its block structure, producing a dense coupling matrix. The resulting network represents a system in which all assets are interconnected, while retaining an intrinsic structure characterized by $M$ zero eigenvalues and rank $K-M$ (Fig. \ref{f4} (b)).

\subsection{Stability analysis}

\begin{figure}[bb]
\centerline{\includegraphics[width=6.0in]{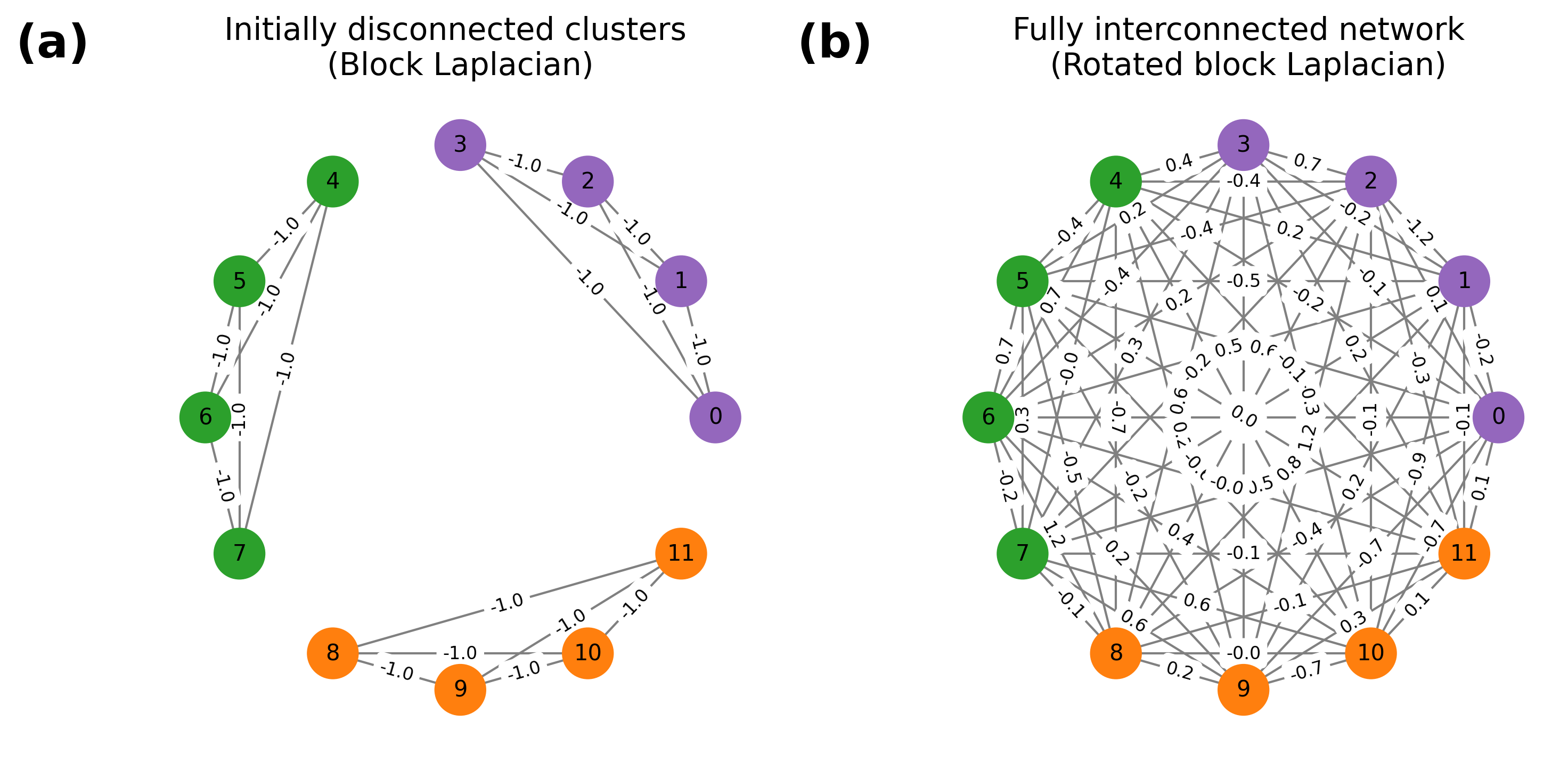}}
\vspace*{8pt}
\caption{Network representation of the coupling structure. (a) Block Laplacian matrix $\mathbf{L}$ composed of $M\ =\ 3$ disconnected clusters, each with $N\ =\ 4$ fully connected components.
(b) Coupling matrix $\mathbf{C}=\mathbf{QLQ}^T$ obtained via orthogonal transformation, preserving the eigenvalue spectrum while producing a fully interconnected network.}
\label{f4}
\end{figure}

Here we link synchronization theory~\cite{pikovsky2001universal} with the emergence of financial factors. We look at the conditions at which the components of a single interconnected cluster (as shown in Fig~\ref{f4} (a)) self-organizes into a synchronized pattern. We interpret the emergence of this pattern as the emergence of a single factor $\mathsf{f}_{t}$. The dynamics of each component in the cluster is determined by this single factor $r^{(n)}_t = \mathsf{f}_{t}$. Note that when the system is fully synchronized there is no residue noise.   

We perform linear stability to a single cluster which dynamics are given by 
\begin{equation}
	r_{t+1}=\left(1-\varepsilon\right)g\left(r_t\right)+\frac{\varepsilon}{N} \mathbf{L_i} g\left(r_t\right).
\label{e_stability}
\end{equation}

\noindent The matrix $\mathbf{L_i}$ has an eigenvalue $\lambda = 0$ with a correspondent eigenvector $v_0 = (1,1,\cdots,1)^T$. All initial conditions projected onto $v_0$
corresponds to the synchronization mode $\mathsf{f}_{t}$~\cite{pikovsky2001universal}. If we introduce the condition $\mathsf{f}_t = r^{(1)}_t = r^{(2)}_t = \cdots = r^{(N)}_t$ into Eq. (\ref{e_stability}) we arrive to the map 
\begin{equation}
	\mathsf{f}_{t+1}=\left(1-\varepsilon\right)g\left(\mathsf{f}_t\right).
\label{e_sync}
\end{equation}

\noindent Typically in synchronization models the system reduces to $\mathsf{f}_{t+1} = g(\mathsf{f}_{t})$~\cite{pikovsky2001universal}, but our system keeps track of the independence each asset has with respect to the interconnected network, which is quantified by the term $1-\varepsilon$. The stability of the synchronization mode, which we interpret as a statistical factor, is determined by its Lyapunov exponent $\eta_f(\varepsilon)$. If $\eta_f(\varepsilon) >0$ the eigenmode associated with the eigenvector $v_0 = (1,1,\cdots,1)^T$ is excited, resulting in a synchronized component in the system dynamics. We calculate the corresponding Lyapunov exponent $\eta_f(\varepsilon)$ using the Benettin algorithm~\cite{benettin1980lyapunov} (Figure 5 (a)).

The rest of the eigenvectors $v_n$ of $\mathbf{L_i}$ are orthogonal to $v_0 = (1,1,\cdots,1)^T$ and their corresponding eigenvalues are $\lambda = -N$. The dynamics projected into these eigenvectors can be interpreted as synchronization errors. Then we can decompose the returns into a synchronization factor and a synchronization error $r_t = \mathsf{f}_t + v_n \mathsf{e}_t$. Note the similarity with Eq. (\ref{e1}) showing that the term $v_n \mathsf{e}_t$ can be interpreted as the noise vector $\xi_t$. In a fully synchronized regime the synchronization error completely disappears and its corresponding Lyapunov exponent $\eta_e < 0$. To get the expression of the error Lyapunov exponent we linearize Eq. (\ref{e_stability}) around $\mathsf{e}_t = 0$ to get to

\begin{equation}
	\mathsf{e}_{t+1}=\Big[(1-\varepsilon) + \frac{\varepsilon}{N} \mathbf{L_i} \Big]\frac{dg\left(\mathsf{f}_t\right)}{d \mathsf{f}_t} \mathsf{e}_{t}
\label{e_lin}
\end{equation}

\noindent Noting that $\mathbf{L_i}\mathsf{e}_{t} = -N \mathsf{e}_{t}$
we get to

\begin{equation}
	\mathsf{e}_{t+1}=\left(1-2\varepsilon\right)\frac{dg\left(\mathsf{f}_t\right)}{d \mathsf{f}_t} \mathsf{e}_{t}
\label{e_sync}
\end{equation}

\noindent The Lyapunov exponent corresponding to these transverse directions is given by

\begin{equation}
	\eta_e(\varepsilon) = Log \big( \Big| \frac{1 - 2 \varepsilon}{1 - \varepsilon} \Big| \big) + \eta_f(\varepsilon).
\label{e_lyae}
\end{equation}

\begin{figure}[bb]
\centerline{\includegraphics[width=5.0in]{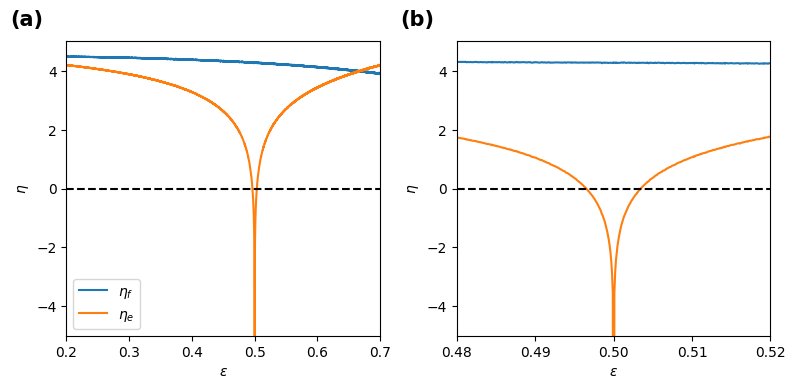}}
\vspace*{8pt}
\caption{Lyapunov exponents governing synchronization behavior.
(a) Synchronization mode exponent $\eta_f$ and the synchronization error exponent $\eta_e$ as function of the coupling parameter $\varepsilon$. 
(b) Zoom around $\varepsilon \approx 0.5$ showing the narrow region where $\eta_e <0$, corresponding to stable synchronization.}
\label{f_lyapunov}
\end{figure}

Figure \ref{f_lyapunov} (a) shows the dependence for both Lyapunov exponents as a function of $\varepsilon$. The Lyapunov exponent for the synchronization mode is always $\eta_f(\varepsilon)>0$, meaning that for the parameter range used here it always exist. This contrast with the error Lyapunov exponent $\eta_e(\varepsilon)$ that is negative in a narrow region around $\varepsilon = 0.5$. Figure \ref{f_lyapunov} (b) shows how narrow is the region where $\eta_e(\varepsilon)<0$, here the system is fully synchronized. We expect that close to the synchronization region and where $\eta_e(\varepsilon)>0$ we will observe a partly synchronized regime consistent with Eq. (\ref{e1}). 

For a system composed by independent clusters and coupled with the full block matrix $\mathbf{L}$, there will be $M$ synchronization modes and $K-M$ error modes. Note that $M$ and $K-M$ correspond with the nullity and the rank of $\mathbf{L}$ respectively. The stability of each of mode is still given by the previous Lyapunov exponents. The coupling matrix $\mathbf{C}$ for the full model (Eq. (\ref{e2})) contains the same eigenvalues as $\mathbf{L}$ but with its eigenvectors pointing in different directions. Therefore we expect that each component in $r_t$ will still have the same number of synchronization modes and error modes, but each synchronization mode (correspoinding to a statistical factor) will weight differently to each component. In the next sections we perform numerical calculations to test this hypothesis. 

\vspace*{0.5cm} 

\section{Results}

\subsection{Statistical Factor Detection and Quantifying the Balance of Loadings}
We identify statistical factors $f_t$ as the principal components of the return vector $r_t$ whose signals exceed a given threshold. In this work, we further investigate how Eq. (\ref{e2}) provided number of factors depend on both the coupling strength $\varepsilon$ and the structural properties of the coupling matrix $\mathbf{C}$. The central question is how many of these components should be considered significant drivers of the system. 

Previous studies addressed this question using random matrix theory (RMT) \citep{de2020machine, laloux2000random, plerou2002random}, where any eigenvalue of the covariance matrix that lies outside the RMT-predicted bounds is deemed significant. To use RMT we need a significant number of assets to obtain a statistical significant eigenvalue distribution, here we employ a method that works even when we look at a small number of assets. To address this, we compare the PCA eigenvalues of the coupled system ($0 < \varepsilon \leq 1$) with that of the uncoupled system ($\varepsilon = 0$), which provides a reference noise baseline.

Taking advantage of the ability to simulate Eq. (\ref{e2}) for the uncoupled system ($\varepsilon=\ 0$), we compare the normalized eigenvalues $\Lambda_k(\varepsilon)$ obtained from PCA between the uncoupled case ($\varepsilon=\ 0$) and a coupled case ($0\ <\ \varepsilon\le1$). The number of significant statistical factors $\hat{M}$ is defined as
\begin{equation}
	\hat{M}=\sum_{k}\theta\left(\Lambda_k(\varepsilon)- \Lambda_k(0)\right) 
\label{e11}
\end{equation}

\noindent where $\theta\left(\cdot\right)$ is the Heaviside step function
\begin{equation}
\theta(x)=
\begin{cases}
1, & x>0, \\ 0, & x\leq 0.
\end{cases}
\label{eheaviside}
\end{equation}

\noindent This definition counts the number of components whose contribution exceeds that observed in the uncoupled system, thereby identifying statistical factors that arise from interactions rather than independent noise.

A straightforward question concerns how factor loadings $\beta_{k,m}$ are distributed (balanced) across assets. For example, when a system is coupled into isolated clusters, as in the graph shown in Fig. \ref{f4} (a), each return $r_t^{\left(k\right)}$ loads exclusively on the factor associated with its own cluster (as mentioned in the stability analysis). In such cases, the factor loadings are clearly imbalanced, with $\beta_{k,m}= 1$ for within clusters and $\beta_{k,m} =0$ for between the others.

To quantify how strongly each asset is influenced by the different factors, we define a measure of loading balance based on the normalized Shannon entropy:
\begin{equation}
	H\left(r^{\left(k\right)}\right)=-\frac{1}{log\left(\hat{M}\right)}\sum_{m=1}^{\hat{M}}{\alpha_{k,m}log\left(\alpha_{k,m}\right)},
\label{e12}
\end{equation}

\noindent where $0\log 0:=0$. The factor weights $\alpha_{k,m}$, measuring the balance of the absolute magnitudes of the loadings, are given by
\begin{equation}
	\alpha_{k,m}=\frac{{|\hat{\beta}}_{k,m}|}{\sum_{m=1}^{\hat{M}}{|\hat{\beta}}_{k,m}|},	 
\label{e13}
\end{equation}

\noindent here factor loadings ${\hat{\beta}}_{k,m}$ are estimated using ordinary least squares (OLS) regression of each asset return $r_t^{(k)}$ on the extracted factor $f_t^{(m)}$:
\begin{equation}
	{\hat{r}}_t^{\left(k\right)}=\sum_{m}{{\hat{\beta}}_{k,m}f_t^{\left(m\right)}}.
\label{e14}	
\end{equation}	

\noindent  If an asset loads predominantly on a single or few factors (biased), the entropy is low; if it loads evenly across factors (balanced), the entropy is high. In particular,  a value of $H\left(r^{\left(k\right)}\right)=1$ indicates perfectly balanced loadings,  while $H\left(r^{\left(k\right)}\right)=0$ corresponds to relations only to a single factor.

\begin{figure}[tb]
\centerline{\includegraphics[width=5in]{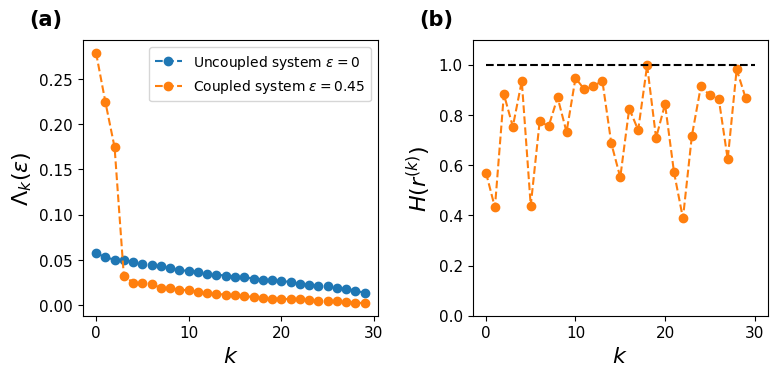}}
\vspace*{8pt}
\caption{Factor identification and loading balance for a single realization (composed of $M = 3$ clusters each with $N= 10$) 
(a) Normalized PCA eigenvalues $\Lambda_k (\varepsilon)$ for the uncoupled system ($\varepsilon=\ 0$) and the coupled system ($\varepsilon = 0.45$), where the uncoupled case defines the noise floor.
(b) Normalized Shannon entropy $H(r_t^{(k)})$ measuring the balance of factor loadings, where values close to 1 indicates well-balanced contributions across factors. }
\label{f5}
\end{figure}

Fig. \ref{f5} illustrates our approach to the questions mentioned above, involving $\mathbf{C}$ for a generated series composed initially of $M=3$ clusters, $K\ =\ 30$, $\varepsilon=\ 0.45$, and 251 points.  Fig. \ref{f5}(a) displays $\Lambda_k(\epsilon)$ for the uncoupled and coupled cases. Fig. \ref{f5} (b) displays $H\left(r^{\left(k\right)}\right)$ values. For this case, most assets exhibit $H_k$ close to 1, and none are equal to zero, showing that the resulting factors are non-trivial and that loadings are generally well balanced. 

\subsection{Emergence of Statistical Factors}

With the definition for the number of significant factors $\hat{M}$ and the normalized Shannon entropy $H\left(r^{\left(k\right)}\right)$ as a measure of the balance of factor weights, we now examine the behavior of the system defined by Eq. (\ref{e2}) as a function of $M$ and $\varepsilon$. Several realizations of Eq. (\ref{e2}) were simulated for $1\le M\ \le6$ in steps of $\Delta M\ =\ 1$, and for $0.2\ \le\varepsilon\le0.7$ in steps of $\Delta\varepsilon=\ 0.01$. The number of elements per cluster was fixed to $N=10$, and for each $(M,\varepsilon)$ pair we performed 200 repetitions.

\begin{figure}[tb]
\centerline{\includegraphics[width=5in]{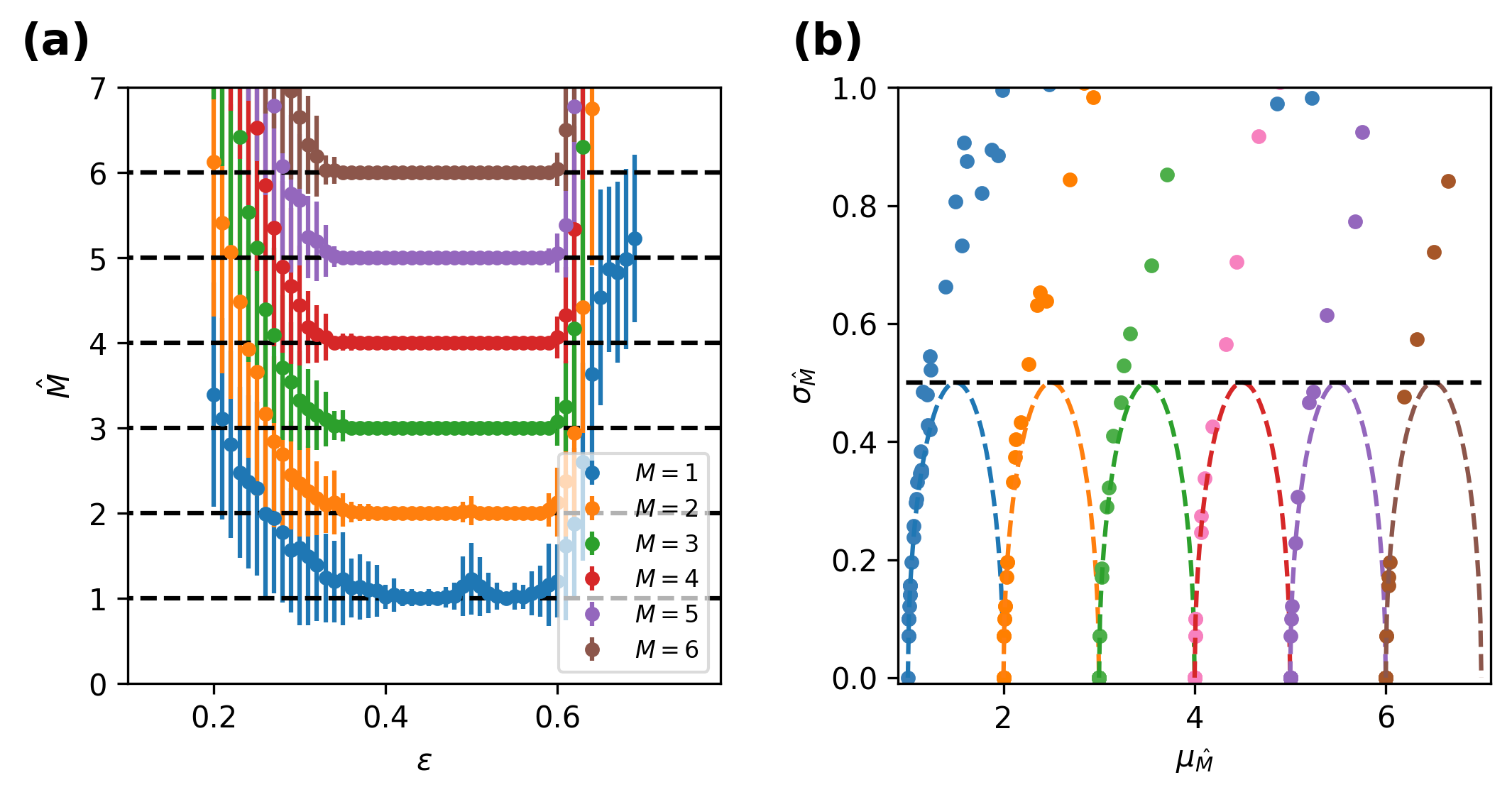}}
\vspace*{8pt}
\caption{Stability of factor detection across simulations.
(a) Estimated number of factors $\hat{M}$ as a function of $\varepsilon$, with error bars computed across ensembles. 
(b) Relationship between the mean $\mu_{\hat{M}}$ and standard deviation $\sigma_{\hat{M}}$, compared with the prediction of a simple binary model (dashed line) where $\hat{M}=M$ is found with probability $p$ and $\hat{M}=M+1$ is found with probability $1-p$. 
Regions where $\sigma_{\hat{M}} \approx 0$ indicate stable and consistent factor detection.}
\label{f6}
\end{figure}

Since we have proposed a method to identify $\hat{M}$, the analysis is performed as a function of the coupling parameter $\varepsilon$. For each iteration, we record the value of $\hat{M}$ and calculate, for each $\varepsilon$, the corresponding sampling mean $\mu_{\hat{M}}$ and the standard deviation $\sigma_{\hat{M}}$. A summary of the results for the number of relevant factors is shown in Fig. \ref{f6}. Panel (a) displays error bars for each $\varepsilon$ to illustrate $\hat{M}$ estimation performance. For most values of $M$ and $\varepsilon$, there is a clear region where the error bars collapse close to zero. In this region, the number of initial clusters $M$ (i.e. the $nullity$ of $\mathbf{C}$) coincides exactly with the number of relevant factors $\hat{M}$. This sharp reduction in variability of the error bar indicates a transition to a regime where the number of detected factors become stable. However, for $M=1$ and $M=2$, we still observe significant error bars in this middle region.

To understand the behavior of observation error bars, consider a simple model in which the number of relevant factors is $\hat{M}=M$ with probability $p$ and $\hat{M}=M+1$ with probability $1-p$. Under this simple binary model, the relationship between the standard deviation $\sigma_{\hat{M}}$ and the mean $\mu_{\hat{M}}$ is given by

\begin{equation}
	\sigma_{\hat{M}}=\sqrt{\left(M-\mu_{\hat{M}}\right)\left(\mu_{\hat{M}}+1-M\right)}.	
\label{e15}
\end{equation}

\noindent Fig. \ref{f6} (b) compares this theoretical relationship (i.e. Eq. (\ref{e15}), dashed line) with $\sigma_{\hat{M}}$ and $\mu_{\hat{M}}$ numerically calculated from the 200 iterations at each value of $\varepsilon$. The agreement is strong up to the maximum value $\sigma_{\hat{M}}=0.5$ given by the model. Thus, cases with $\sigma_{\hat{M}}<0.5$ in Fig. \ref{f6} (a) can be interpreted as parameter regions where either $\hat{M}=M$ or $\hat{M}=M+1$ is detected.

\begin{figure}[tb]
\centerline{\includegraphics[width=5in]{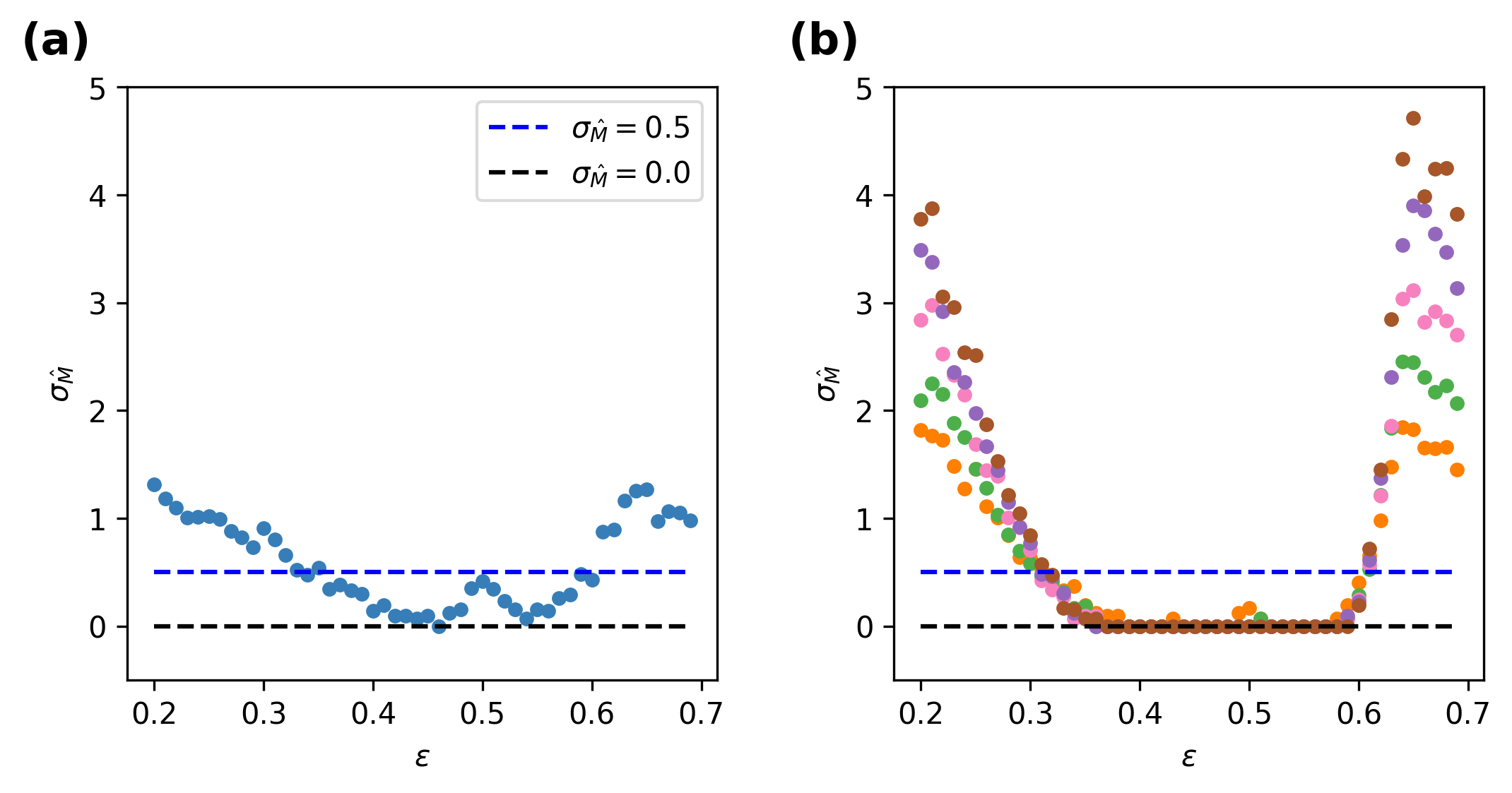}}
\vspace*{8pt}
\caption{Variability in the detected number of factors $\hat{M}$ as a function of $\varepsilon$. 
(a) Case $M=1$, where $\sigma_{\hat{M}}$ varies significantly across $\epsilon$. However, in the region $0.35\ <\ \varepsilon\le0.6$ the standard deviation fluctuates only between 0.0 and 0.5. 
(b) Cases $M=2$--$6$, showing a stable region $0.36\ <\ \varepsilon\le0.58$ where $\sigma_{\hat{M}}=0$, indicating consistent factor emergence.}
\label{f7}
\end{figure}

As mentioned above, when we observe M synchronization modes, each asset may follow a common factor $f_t^{\left(m\right)}$ related to its initial cluster in the Laplacian
matrix $\mathbf{L}$. To examine this possibility, we evaluated the mean $\mu_H$ and standard deviation $\sigma_H$ of the normalized Shannon entropy. Again, the normalized Shannon entropy in Eq. (\ref{e12}), provides a measure of how balanced the factor loadings $\beta_{k,m}$ are for each return $r_t^{\left(k\right)}$. Fig. \ref{f7} illustrates how factors indeed emerge within a specific region of the parameter space, namely $0.36 < \epsilon \leq 0.58$. For $M =1$ this emergence is not consistent (Fig. \ref{f7} (a)), whereas for $M = 2$ through 6, the behavior is highly consistent across simulations (Fig. \ref{f7} (b)).

\subsection{Balance of Factor Weights}

Fig. \ref{f8} (a) shows that in the same region where $\sigma_{\hat{M}}$ is close to zero, the mean Shannon entropy is reduced to a non-zero value. Moreover, the mean entropy $\mu_H$ increases with $M$: systems with a large number of clusters exhibit higher average entropy. At the same time, as the standard deviation $\sigma_H$ also increases, but its magnitude decreases as a function of $M$. For example, variations are approximately $25\%$ for $M=2$, while for $M=6$ they decrease to about $7.5\%$ (Fig. \ref{f8} (b)). Therefore, these observations conclude that the factor loadings in this model are well balanced.

\subsection{Cumulative Explained Variance Ratio}

We examine the fluctuations around the synchronization manifold in phase space. This manifold, spanned by $\hat{M}$ synchronization modes (identified empirically as statistical factors), attracts nearby trajectories. However, the existence of these synchronization modes does not imply that all fluctuations orthogonal to the manifold have completely disappeared. To quantify how strongly the dynamics are concentrated around this $\hat{M}$-dimensional manifold, we compute the cumulative explained variance ratio associated with the detected statistical factors,$\sigma_f^2(\varepsilon)$. Because the eigenvalues $\Lambda_k(\varepsilon)$ are normalized, the resulting quantity measures the proportion of the total variance of the return vector $r_t$ captured by the leading statistical factors (e.g., the three initial values for $\varepsilon=\ 0.45$ in Fig. \ref{f5} (b) ). The cumulative explained variance ratio is given by

\begin{equation}
	\sigma_f^2 (\varepsilon) =\sum_{k=1}^{\hat{M}}\Lambda_k(\varepsilon)	
\label{e16}
\end{equation}

\begin{figure}[t]
\centerline{\includegraphics[width=6.5in]{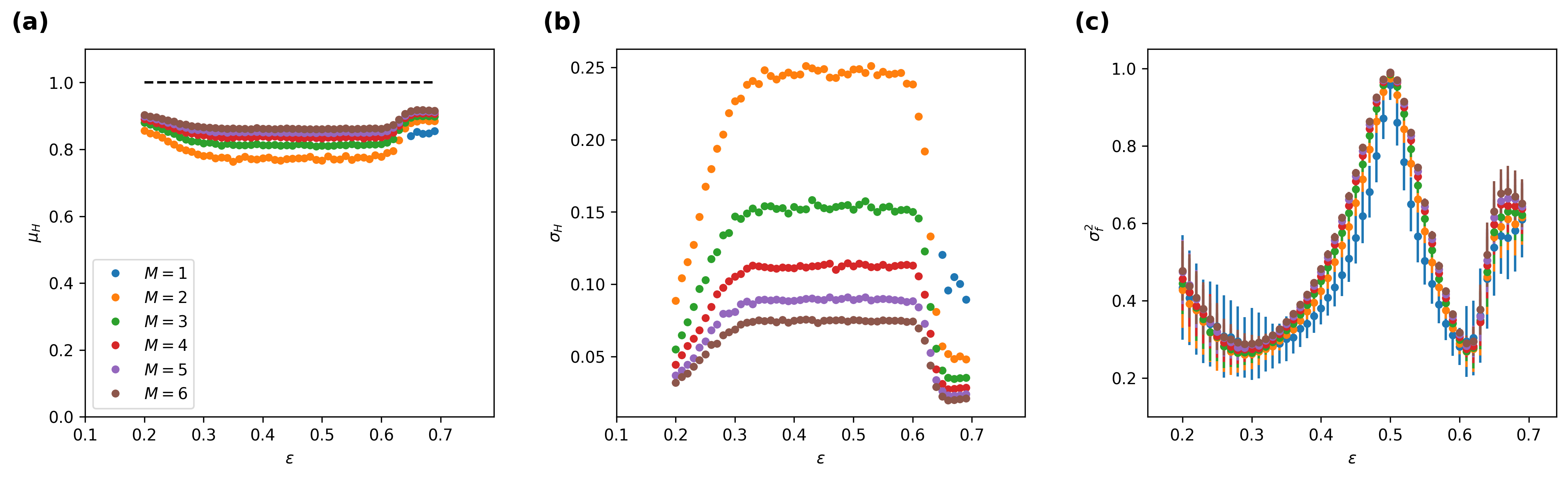}}
\vspace*{8pt}
\caption{Properties of factor loadings and explained variance. (a) Mean normalized Shannon entropy $\mu_H$ indicating average balance of factor loadings.
(b) Standard deviation $\sigma_H$ showing variability across assets. However, this is capped for all combinations of $\varepsilon$ and $M$.
(c) Cumulative explained variance $\sigma_f$ attributed to the detected factors.
Near $\varepsilon \approx 0.5$, the leading factors explain nearly all of the variance, indicating effective dimensionality reduction.}
\label{f8}
\end{figure}

Fig. \ref{f8} (c) shows that the cumulative explained variance approaches unity near $\varepsilon=\ 0.5$ (as expected from the stability analysis). This indicates that the statistical factors capture nearly all of the variance of the return vector $r_t$, implying that the simulated dynamics are effectively concentrated around a low-dimensional synchronization manifold. Thus, close to $\varepsilon = 0.5$, there exists an attracting region in phase space where the leading statistical factors account most of the variance of the system.  The remaining proportion $1-\sigma_f^2(\varepsilon)$ corresponds to variance not captured by the detected statistical factors. In the theoretical framework developed in the Stability Analysis section,  fluctuations orthogonal to the synchronization manifold are represented by synchronization error modes. Therefore, a decrease in $1-\sigma_f^2(\varepsilon)$ can be interpreted as a reduction in the activity of these error modes and, consequently, as a stronger confinement of the dynamics to the synchronization manifold. However, this interpretation should be viewed as a model-based explanation of the residual variance rather than a direct measurement of noise. Consequently, a cumulative explained variance ratio close to unity indicates that the detected statistical factors provide an increasingly complete description of the system dynamics, while not implying that all residual fluctuations vanish.

\subsection{Empirical Observations}

To illustrate these mechanisms, we tested the output of our simulations against empirical small portfolio. The portfolio is composed of three indexes and seven stocks (see Table \ref{tab:tickers}). Two different time periods were analyzed, the first one from 2020-07-25 to 2021-07-25 and the second one from 2021-07-25 to 2022-07-25. From the shape of the PCA normalized eigenvalue spectrum we conjectured that there are two significant empirical statistical factors in the first period and a single significant factor for the second one. We proceeded by performing numerical simulations setting the values to $M=2$ for the first period and $M=1$ for the second one. In both cases we did a numerical sweeps $0 \leq \varepsilon \leq 1$ and keeping the random seed fixed (see the \href{https://colab.research.google.com/drive/1NrQKD1ibd9M9t3S5f7-7BoHWocIkz2aO?usp=sharing}{jupyter notebook} for outputs with different seeds). The aim was to find an eigenvalue spectrum that resembles the empirical one, this was determined by quantifying the error between them. 

\begin{figure}[t]
\centerline{\includegraphics[width=6.5in]{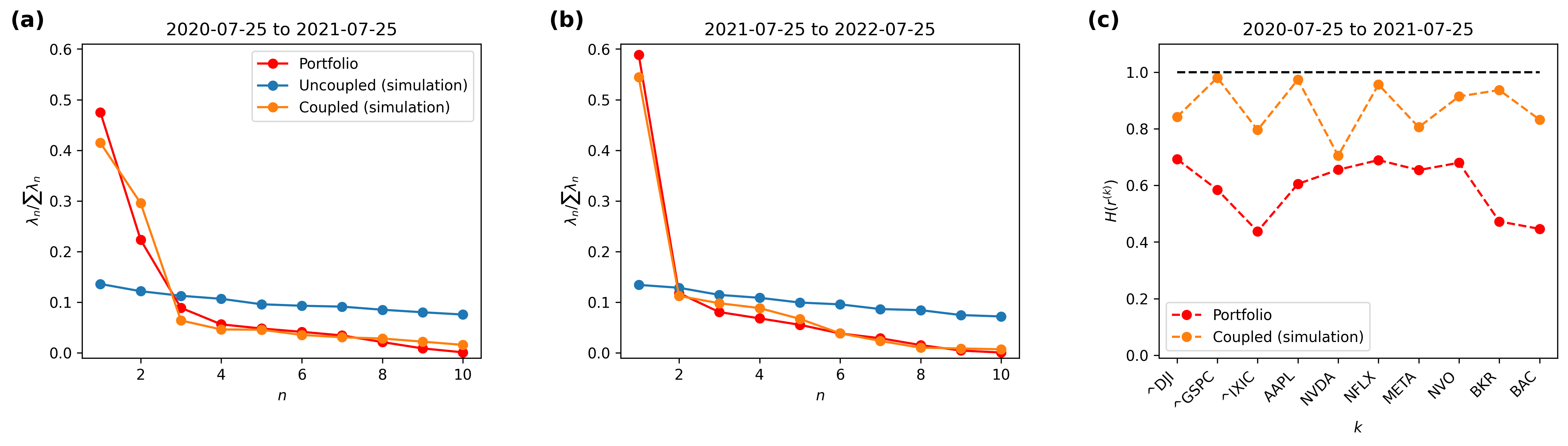}}
\vspace*{8pt}
\caption{Comparison between simulated and empirical factor structure.
(a)--(b) PCA eigenvalue spectra for empirical portfolio data and numerical simulations for the periods 2020-07-25 to 2021-07-25 and 2021-07-25 to 2022-07-25, with the uncoupled case defining the noise floor. The model reproduces qualitative features observed in real data. 
(c) Comparison of normalized Shannon entropy between simulated data and empirical portfolio returns. The entropy comparison shows that empirical factors are less balanced.}
\label{f9}
\end{figure}

In Figure \ref{f9} (a) and (b) we show the portfolio eigenvalue spectrum, along with the spectrum obtained numerically for both the coupled and uncoupled case. In these examples, we found the minimum error at  $\varepsilon = 0.56$ and for the second case  at  $\varepsilon = 0.46$. This is relatively close to the optimum of $\varepsilon = 0.5$. Note that the shape of the empirical and simulated spectrum resembles each other, specially in initially having a fast drop followed by a slow decrease. As in previous sections, the uncoupled case is used as a reference to differentiate between factors and the noise floor. Two empirical statistical factors are above the noise floor in the first period and one for the second one as we conjectured beforehand. Interestingly, the synchronization structure evolved over time, with returns initially exhibiting two distinct synchronization modes that later gave way to a single dominant synchronization mode.

Finally, we compared the normalized Shannon entropy between the numerical run and for the first period empirical portfolio  (Fig. \ref{f9} (c)). As expected from before (Figure \ref{f8} (a)), the normalized Shannon entropy for the simulation fluctuates around an average of $H = 0.87$, meaning that factor weights tend to be well balanced. The entropy for the empirical portfolio fluctuates around an average of $H= 0.59$, meaning that factor weights tend to be not balanced, yet they are not completely described by a single factor $(H=0)$. This shows that in the empirical portfolio each asset tends to be biased towards a given factor, still the remaining one plays a significant role. This is specially surprising for the indices $\hat{}DJI$, $\hat{}GSPC$ and $\hat{}IXIC$ as they tend to be proxies of the market factor\cite{avellaneda2022principal}, and for this time period they are described by two factors. Altogether the resemblance between numerical and empirical data provide an illustrative example of how statistical factors could arise from the coupling between assets, although a formal empirical validation is beyond the scope of the present study.

\begin{table}[h]
\centering
\begin{tabular}{l|l}
\hline
Ticker & Company / Index Name \\
\hline
\textasciicircum DJI \   & \ Dow Jones Industrial Average \\
\textasciicircum GSPC \  & \ S\&P 500 Index \\
\textasciicircum IXIC \  & \ Nasdaq Composite Index \\
AAPL \   & \ Apple Inc. \\
NVDA \   & \ NVIDIA Corporation \\
NFLX \   & \ Netflix, Inc. \\
META \   & \ Meta Platforms, Inc. \\
NVO \    & \ Novo Nordisk A/S \\
BKR \    & \ Baker Hughes Company \\
BAC \    & \ Bank of America Corporation \\
\hline
\end{tabular}
\caption{List of tickers and corresponding names used in the empirical portfolio in Figure \ref{f9}.}
\label{tab:tickers}
\end{table}

\section{Discussion and Conclusion}

The model demonstrates that statistical factors can emerge from a diffusion process (i.e. interactions) among assets, rather than being imposed externally. The dimensionality reduction observed in the system aligns with the emergence of synchronization modes, where the nullity of the coupling matrix determines the number of dominant modes. These patterns serve as latent dimensions of the system, analogous to factors in standard financial models. 

Our numerical findings highlight several key results. First, for $M>2$ the number of factors that emerge in the critical region $(0.36\ <\ \varepsilon\le 0.58)$ coincides with the number of initial clusters, or equivalently, with the $nullity$ of the coupling matrix $\textbf{C}$. This suggests that the emergence of factors is governed by the emergence of synchronization modes. We identify an optimal coupling strength $\varepsilon$ via stability analysis, at which the statistical factors account for nearly all of the variance of the return system, leaving only a small residual contribution associated with deviations from the synchronization manifold. Finally, we showed that the model reproduces features of the PCA eigenvalue spectrum for a small portfolio. The techniques developed in this paper allow us to recognize between the noise floor and the signal. Interestingly in one time period our portfolio is described by two factors, and in another by a single factor. This suggests that the connectivity between their interaction networks changed such that one synchronization mode disappeared in the second case. Taken together this work shows that statistical, and potentially financial factors, might correspond to different synchronization modes that emerge via network interactions.

\section*{Contributions Statement}
J. Negrete Jr. conceived the study. All authors performed calculations, developed R/Python code, analyzed simulations, and wrote the manuscript. All authors discussed the results and contributed to the final manuscript.

\section*{Data Access Statement}
Data supporting the findings of this study are openly available in Yahoo Finance, retrieved using Python's package \verb|yfinance|. The Jupyter notebook used to perform the analysis and figures of this paper is available in Google Colab \url{https://colab.research.google.com/drive/1NrQKD1ibd9M9t3S5f7-7BoHWocIkz2aO?usp=sharing}.

\bibliography{export}

\end{document}